\documentclass[%
 aip,
 amsmath,amssymb,
 reprint,%
]{revtex4-2}

\usepackage{graphicx}
\usepackage{dcolumn}
\usepackage{bm}

\usepackage[utf8]{inputenc}
\usepackage[T1]{fontenc}
\usepackage{mathptmx}

\begin{document}

\preprint{AIP/123-QED}

\title[A growth diagram for chemical beam epitaxy of GaP$_{1-x}$N$_{x}$ alloys on nominally $(001)$-oriented GaP-on-Si substrates]{A growth diagram for chemical beam epitaxy of GaP$_{1-x}$N$_{x}$ alloys on nominally $(001)$-oriented GaP-on-Si substrates}

\author{K. Ben~Saddik}
\email{karim.bensaddik@estudiante.uam.es (corresponding author)}
\affiliation{Electronics and Semiconductors Group (ElySe), Applied Physics Department, Universidad Autónoma de Madrid,
ES-28049 Madrid, Spain}%

\author{B. J. García}%
\email{basilio.javier.garcia@uam.es (main corresponding author)}
\affiliation{Electronics and Semiconductors Group (ElySe), Applied Physics Department, Universidad Autónoma de Madrid,
ES-28049 Madrid, Spain}%
\affiliation{Instituto Nicolás Cabrera, Universidad Autónoma de Madrid, ES-28049, Madrid, Spain}

\author{S. Fernández-Garrido}
\email{sergio.fernandezg@uam.es (main corresponding author)}
\affiliation{Electronics and Semiconductors Group (ElySe), Applied Physics Department, Universidad Autónoma de Madrid,
ES-28049 Madrid, Spain}%
\affiliation{Instituto Nicolás Cabrera, Universidad Autónoma de Madrid, ES-28049, Madrid, Spain}

\date{\today}

\begin{abstract}
The compound GaP$_{1-x}$N$_{x}$ is highly attractive to pseudomorphically integrate red light-emitting devices and photovoltaic cells with the standard Si technology because it is lattice matched to Si with a direct band gap of $\approx1.96$~eV for $x=0.021$. Here, we report on the chemical beam epitaxy of GaP$_{1-x}$N$_{x}$ alloys on nominally $(001)$-oriented GaP-on-Si substrates. The incorporation of N into GaP$_{1-x}$N$_{x}$ was systematically investigated as a function of the growth temperature and the fluxes of the N and P precursors, 1,1-dimethylhydrazine (DMHy) and tertiarybutylphosphine (TBP), respectively. We found that the N mole fraction exhibits an Arrhenius behavior characterized by an activation energy of $(0.79\pm 0.05)$~eV. With respect to the fluxes, we determined that the N mole fraction is linearly proportional to the flux of DMHy, and inversely proportional to the one of TBP. All results are summarized in a universal equation that describes the dependence of \textit{x} on the growth temperature and the fluxes of the group-V precursors. The results are further illustrated in a growth diagram that visualizes the variation of \textit{x} as the growth temperature and the flux of DMHy are varied. This diagram also shows how to obtain single-phase and flat GaP$_{1-x}$N$_{x}$ layers, as certain growth conditions result in chemically phase-separated layers with rough surface morphologies. Last, our results demonstrate the feasibility of chemical beam epitaxy to obtain single-phase and flat GaP$_{1-x}$N$_{x}$ layers with \textit{x} up to about $0.04$, a value well above the one required for the lattice-matched integration of GaP$_{1-x}$N$_{x}$-based devices on Si.
\end{abstract}

\maketitle

\section{\label{sec:level1}Introduction}
The incorporation of small amounts of N into GaP, besides decreasing the lattice parameter, induces an indirect-to-direct band gap transition well described by the band anticrossing model \cite{Shan2004,Chamings2009}. Remarkably, for a N mole fraction $x=0.021$, the ternary compound GaP$_{1-x}$N$_{x}$ is lattice-matched to Si with a direct band gap of about $1.96$~eV at room temperature, making this material rather unique for the monolithic integration of pseudomorphic red-light emitters and III-V photovoltaic solar cells with the widespread, highly scalable and cost-effective Si technology \cite{Xin2000a,Cornet2011,Robert2012,Jain2014,Durand2014,Sukrittanon2015,Kharel2018,BenSaddik2019}. Nevertheless, despite of the great potential of this compound, commercial red-light emitting devices are still based on Al$_{x}$In$_{y}$Ga$_{1-x-y}$P alloys \cite{Tsao2014} and the efficiency of GaP$_{1-x}$N on Si photovoltaic solar cells remains too low as to consider this material combination a competitive technology \cite{Sukrittanon2015}. This situation arises from: (i) The challenging synthesis of GaP$_{1-x}$N alloys with high structural perfection. Specifically, the material quality of dilute-nitride compounds is known to degrade as the N content and the layer thickness are increased, as reported for thin films and devices grown, for instance, by metal-organic vapor phase epitaxy (MOVPE) and plasma-assisted molecular beam epitaxy (PA-MBE) \cite{Xin2000a,Yonezu2002,Almosni2013,Tilli2014,Yamane2018,Balanta2020}. (ii) The common use of highly misoriented ($4-6$~$^{\circ}$) instead of on-axis Si$(001)$ substrates to avoid the creation of anti-phase domains,\cite{Yonezu2002,Nagarajan2013,Li2017} a solution that undermines the integration of\break GaP$_{1-x}$N$_{x}$-based devices with the standards of Si technology. 

Chemical beam epitaxy (CBE), an ultra-high-vacuum epitaxial growth technique characterized by the use of gaseous precursors in the form of molecular beams for both group-III and -V elements \cite{Houng1992}, offers important advantages over alternative methods for the synthesis of GaP$_{1-x}$N$_{x}$ compounds. On the one hand, the ultra-high-vacuum environment enables the use of powerful techniques to optimize and monitor the growth, such as reflection high-energy electron diffraction (RHEED). On the other hand, as N is not produced by a radio-frequency N$_{2}$ plasma source, there is no crystal-damage associated to the impingement of energetic ionized N species \cite{MiguelSanchez2004}, a major issue in PA-MBE that enforces the realization of ex-situ rapid-thermal annealing treatments to improve the material quality \cite{Utsumi2003,MiguelSanchez2004,Moon2004,Balanta2020}. Furthermore, from technical and safety points of view, CBE is also advantageous as compared to PA-MBE because the residual gases can be easily extracted during and after the growth process. Therefore, it is possible to avoid the risks associated to the accumulation of hazardous P-containing materials inside the epitaxial reactor. So far, however, reports on the synthesis of GaP$_{1-x}$N$_{x}$ compounds by CBE are scarce, limited to the analysis of individual growth parameters on the chemical composition, and restricted to the use of either bulk GaP$(001)$ or misoriented Si$(001)$ substrates \cite{Li1996,Suemune2002,BenSaddik2019}. There is thus the need of comprehensively exploring the chemical beam epitaxy of GaP$_{1-x}$N$_{x}$ alloys on nominally $(001)$-oriented Si substrates to elucidate the potential of CBE for monotonically integrating GaP$_{1-x}$N$_{x}$ devices on Si. 

In this work, we investigate the chemical beam epitaxy of GaP$_{1-x}$N$_{x}$ alloys on CMOS-compatible GaP on Si$(001)$ substrates, as desired for the integration of III-V compounds on Si, with the aim of establishing how to control the chemical composition and properties of these compounds in CBE. Upon independently analyzing the influence of the growth temperature and the fluxes of the group-V precursors on the incorporation of N into GaP$_{1-x}$N$_{x}$, all results are summarized in a universal equation that we used to construct a growth diagram. This diagram, which illustrates the dependence of the chemical composition on the growth conditions as well as the impact of the growth parameters on both the chemical homogeneity and the surface morphology, can be used as a guide to control the properties of GaP$_{1-x}$N$_{x}$ compounds grown by CBE. Finally, on the basis of the studies presented here, we conclude on the feasibility of CBE to produce chemically homogeneous and flat GaP$_{1-x}$N$_{x}$ layers lattice matched to Si. 

\section{Experimental}
\begin{figure}
\includegraphics*[width=0.48\textwidth]{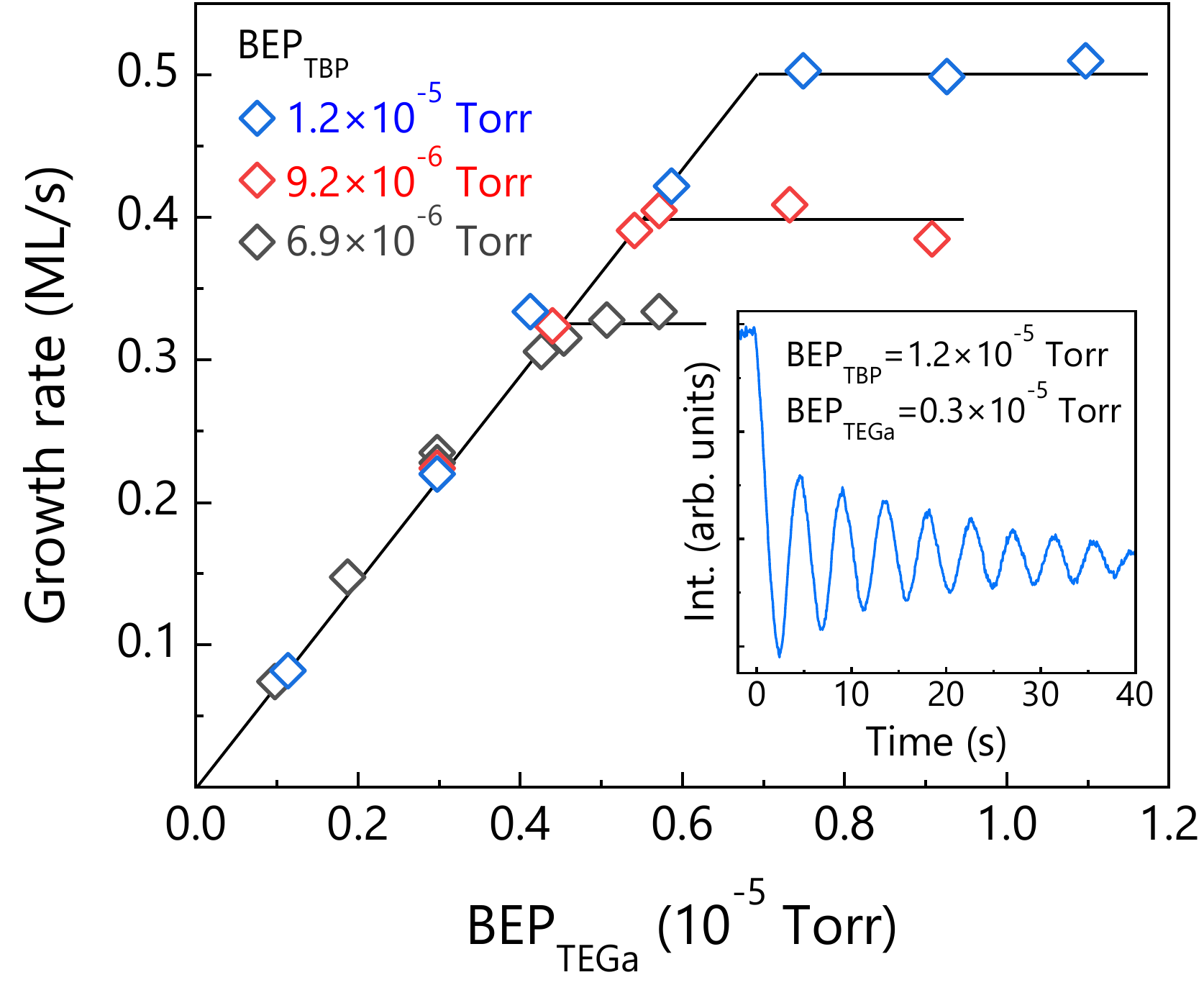}
  \caption{\label{Figure1} GaP growth rate in ML/S, derived from the frequency analysis of RHEED intensity oscillations at 580~$^{\circ}$C, as a function of BEP$_{TEGa}$ for three different values of BEP$_{TBP}$, as indicated in the figure. The solid lines are linear fits to the experimental data within the Ga- and P-limited growth regimes. The inset shows, as an example, the initial temporal evolution of the RHEED intensity for the growth conditions employed in this work to prepare the GaP buffer layers.}
\end{figure}

We prepared the samples in a Riber CBE32 system using triethylgallium (TEGa), tertiarybutylphosphine (TBP) and 1,1-dimethylhydrazine (DMHy) as gas sources for Ga, P and N, respectively. Low-temperature (120~$^{\circ}$C) gas injectors were used for both TEGa and DMHy, while a high-temperature one (820~$^{\circ}$C) was employed for TBP to increase its cracking efficiency, as attested by quadrupole mass spectrometry. The growth chamber is equipped with an infrared optical pyrometer for monitoring the substrate temperature, a 15~keV \break RHEED gun, and an ion gauge that can be placed at the position of the substrate holder during growth to assess the beam equivalent pressure (BEP) of each gas precursor, a magnitude directly proportional to its flux. In the cases of Ga and P, we previously correlated the BEP values of the gas precursors with their actual effective fluxes (i.e., the amount of material that is actually incorporated per time and area units) in equivalent growth rate units of monolayers per second (ML/s) by measuring the growth rate of GaP for different values of BEP$_{TEGa}$ and BEP$_{TBP}$. We note that 1 ML corresponds to $2/a_{Si}^{2}$, where $a_{Si}$ is the Si lattice constant, i.\,e., 1~ML~$=6.78\times 10^{14}$~ atoms/cm$^{2}$. The correlation between the BEP values and the fluxes is illustrated in Fig.~\ref{Figure1}, where the growth rate of GaP, extracted from the frequency analysis of RHEED intensity oscillations, is represented as a function of BEP$_{TEGa}$ for three different values of BEP$_{TBP}$. As can be observed, the growth rate linearly increases with BEP$_{TEGa}$ (P-rich regime) until becoming limited by BEP$_{TBP}$ (Ga-rich regime). Further technical details about the CBE system can be found elsewhere \cite{AitLhouss1994}. 

All samples were grown on $2\times2$~cm$^{2}$ GaP-on-Si$(001)$ substrates diced from a 12" wafer purchased from\break $NAsP_{III/V}$. This wafer, despite its nominally exact $(001)$ orientation, exhibits a slight miscut of $\approx 0.3^{\circ}$ towards one of the four $\left< 110 \right>$ directions, as determined by high-resolution X-ray diffraction (HRXRD). The GaP layer, free of dislocations, stacking faults and twins, is 25~nm thick and possess a smooth surface (root mean square roughness of 0.3~nm) without anti-phase domains. The substrates were In-bonded onto Mo holders and outgassed under TBP supply inside the growth chamber at 610~$^{\circ}$C until observing the appearance of a clear $(2\times4)$ surface reconstruction. The latter reflects the complete thermal desorption of the native oxide layer formed on the purchased wafer due to air exposure. Upon desorbing the oxide, we prepared a 15~nm thick GaP buffer layer at 580~$^{\circ}$C to planarize the surface and bury possible impurities. This layer was grown under P-rich growth conditions with BEP$_{TBP}=1.2\times10^{-5}$~Torr and BEP$_{TEGa}=0.3\times10^{-5}$~Torr, values that correspond to fluxes of 0.5 and 0.22~ML/s, respectively (see Fig.~\ref{Figure1}). Under these conditions GaP grows layer-by-layer, as evidenced by the observation of strong RHEED intensity oscillations (see inset in Fig.~\ref{Figure1}). After the growth of the GaP buffer layer, GaP$_{1-x}$N$_{x}$ was grown for $60$~min, which results in a thickness of about\break 210~nm. As discussed below, different series of samples were prepared to independently investigate the impact of the growth temperature and the fluxes of DMHy and TBP on N incorporation taking as starting point the V/III ratio employed for the growth of the GaP buffer layer, i.\,e., $2.3$.

\begin{figure*}
\includegraphics*[width=0.85\textwidth]{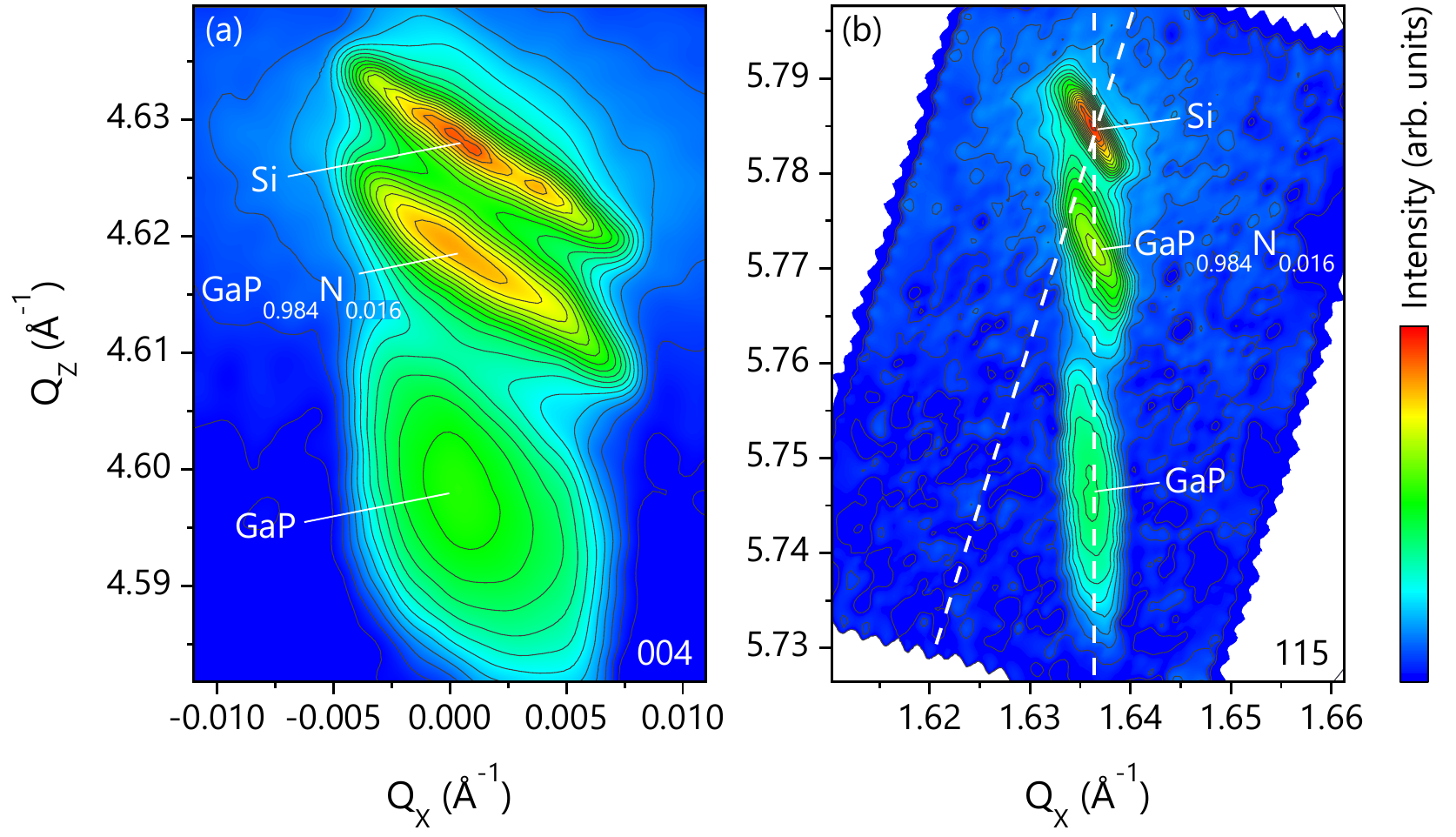}
  \caption{\label{Figure2} Exemplary RSMs around (a) the symmetrical $004$ and (b) the asymmetrical $115$ Bragg reflections of GaP$_{1-x}$N$_{x}$, GaP and Si for a sample grown at 500~$^{\circ}$C with BEP$_{DMHy}=2.4\times10^{-5}$~Torr and BEP$_{TBP}=1.2\times10^{-5}$~Torr. The $115$ RSM was acquired under grazing exit. The vertical and inclined dashed lines in (b) indicate the (Q$_{x}$,Q$_{z}$) coordinates corresponding to fully strained and fully relaxed epilayers on Si, respectively. For the sample shown here with $x=0.016$, both the GaP$_{1-x}$N$_{x}$ and GaP layers are fully strained on the Si substrate. The X-ray intensity is represented in arbitrary units using a logarithmic color scale, as indicated by the scale bar shown on the right.}
\end{figure*}

High-resolution X-ray diffraction measurements were performed with CuK$_{\alpha1}$ radiation (wavelength $\lambda=1.54056$~\r{A}) using a BEDE D3 diffractometer equipped with a 4-bounce Si$(220)$ double-crystal monochromator and a Göbel mirror. Both reciprocal space maps (RSMs) and $\omega /2\theta$ scans were acquired with a $0.5$~mm slit in front of the detector. All measurements were performed along the direction of the miscut. The chemical composition and the strain state of the layers were deduced from the peak locations in RSMs recorded along symmetrical $004$ and asymmetrical $115$ Bragg reflections. As can be observed in the exemplary RSMs shown in Fig.~\ref{Figure2}, we clearly resolve the GaP$_{1-x}$N$_{x}$, GaP and Si reflections. To calculate the chemical composition and the strain, we took into account the possible tilt of the epitaxial layers with respect to the normal of the Si substrate surface, $\gamma$, as described in Refs.~\citenum{Roesener2013,Chauveau2003}. The obtained values of $\gamma$ are $\approx 10$~arcsec, i.\,e., one order of magnitude smaller than those reported for GaP layers grown on misoriented Si$(001)$ substrates \cite{Takagi2010}. The GaP$_{1-x}$N$_{x}$ stiffness constants were estimated according to the Vegard's law. The GaP and GaN stiffness constants and lattice parameters were taken from Ref.~\citenum{Vurgaftman2001}.  

\section{Results and discussion}
\subsection{\label{section3.1}Effect of growth parameters on N incorporation}
\begin{figure*}
\includegraphics*[width=\textwidth]{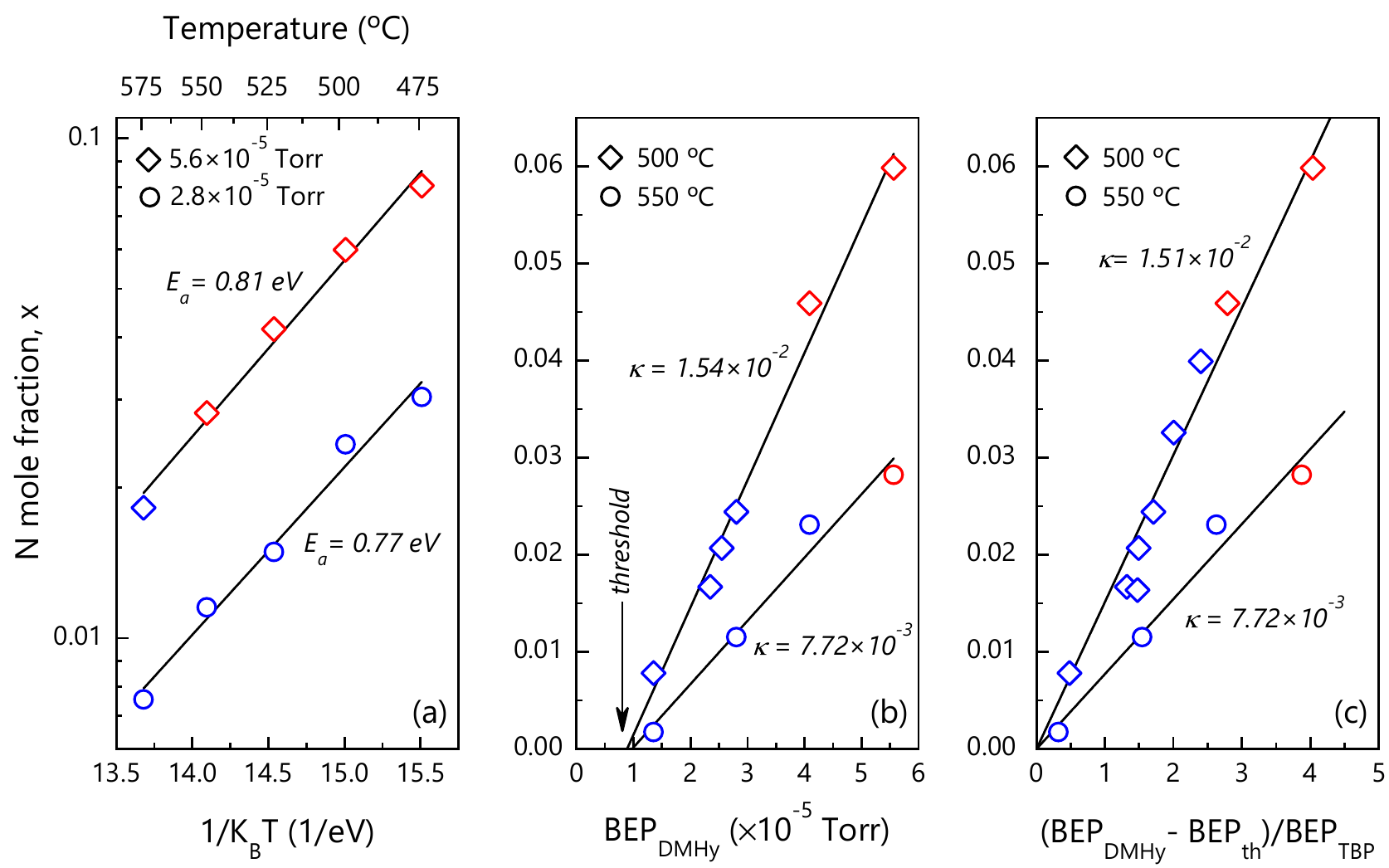}
  \caption{\label{Figure3}(a) N mole fraction \textit{x} as a function of the growth temperature T for a BEP$_{TBP}$ of $1.2\times10^{-5}$~Torr and two different BEP$_{DMHy}$ values, as indicated in the figure; the solid lines represent Arrhenius fits to the experimental data yielding activation energies of $E_{a}=0.81\pm0.06$~eV for BEP$_{DMHy}=5.6\times10^{-5}$~Torr, and $E_{a}=0.77\pm0.05$~eV for BEP$_{DMHy}=2.8\times10^{-5}$~Torr. (b) N mole fraction \textit{x} as a function of BEP$_{DMHy}$ at two different growth temperatures, 500 and 550~$^{\circ}$C, for BEP$_{TBP}=1.2\times10^{-5}$~Torr; the solid lines are linear fits to the experimental data yielding $\kappa=(1.54\pm 0.09)\times10^{-2}$ for 500~$^{\circ}$C and $\kappa=(7.72\pm 0.9)\times10^{-3}$ for 550~$^{\circ}$C. (c) N mole fraction \textit{x} as a function of (BEP$_{DMHy}$-BEP$_{offset}$)/BEP$_{TBP}$ at 500 and 550~$^{\circ}$C. The linear fits to the experimental data shown by the solid lines yield $\kappa$ values of $(1.51\pm0.05)\times10^{-2}$ and $(7.7\pm0.4)\times10^{-3}$ for 500 and 550~$^{\circ}$, respectively. In all cases, blue (red) data points represent samples without (with) chemical phase separation according to HRXRD measurements. For phase-separated samples, the N mole fraction represents the mean value.}
\end{figure*}
In this section, we systematically analyze the individual impact of the growth temperature and the fluxes of DMHy and TBP on the incorporation of N into GaP$_{1-x}$N$_{x}$. Figure~\ref{Figure3}(a) shows the N mole fraction \textit{x} as a function of the growth temperature for BEP$_{TBP}=1.2\times10^{-5}$~Torr and two different values of BEP$_{DMHy}$, namely, $2.8\times10^{-5}$ and $5.6\times10^{-5}$~Torr. The N mole fractions shown in the figure are mean values, as some samples exhibit evidences of chemical phase separation (\textit{x} values of phase-separated samples are shown using red color in Fig.~\ref{Figure3}). We will return to this point in section~\ref{section1}. Regardless of the BEP$_{DMHy}$ value, we observe an exponential decrease of \textit{x} with increasing growth temperature. Specifically, as the growth temperature is varied between 475 and 575~$^{\circ}$C, \textit{x} decreases from 0.03 to 0.008 for BEP$_{DMHy}=2.8\times10^{-5}$~Torr, and from 0.081 to 0.018 for BEP$_{DMHy}=5.6\times10^{-5}$~Torr. The maximum N mole fraction achieved here using DMHy as N source, 0.08, is significantly higher than the previous values reported for GaP$_{1-x}$N$_{x}$ layers grown by CBE using either NH$_{3}$\cite{Li1996} or monomethylhydrazine (MMHy)\cite{Suemune2002} ($x=0.0008$ for Ref.\citenum{Li1996} and 0.0189 for Ref.\citenum{Suemune2002}). Returning to Fig.~\ref{Figure3}(a), the observed temperature dependencies of \textit{x} are well described by Arrhenius laws. The fits to the experimental data included in the above figure yield activation energies of $(0.77\pm0.05)$ and $(0.81\pm0.05)$~eV for BEP$_{DMHy}=2.8\times10^{-5}$ and $5.6\times10^{-5}$~Torr, respectively. The activation energies are thus rather similar when taking into account the experimental uncertainties. Hence, the incorporation of N as a function of the growth temperature can be written as:
\begin{equation}
x = A~exp~(E_{a}/k_{B}T), \label{eqn:Arrhenius}
\end{equation}
where $k_{B}$ is the Boltzmann's constant, by using the mean activation energy value of E$_{a}=(0.79\pm 0.05)$~eV and a BEP$_{DMHy}$ dependent exponential prefactor \textit{A}. In the present case, we found \textit{A}$=2.05\times10^{-7}$ for BEP$_{DMHy}=2.8\times10^{-5}$~Torr and \textit{A}$=2.97\times10^{-7}$ for BEP$_{DMHy}=5.6\times10^{-5}$~Torr. The decrease of \textit{x} with the growth temperature is comparable to those reported for samples grown by plasma-assisted gas-source molecular beam epitaxy (GS-MBE) \cite{Bi1996} and MOVPE using DMHy \cite{Kunert2004,Wakahara2007}. In the case of MOVPE, the temperature dependence of \textit{x} was also well described by Arrhenius laws, but with slightly higher activation energies, specifically, $0.94-1$~eV \cite{Kunert2004,Wakahara2007}. In accordance to previous reports, we attribute the decreasing N content with increasing growth temperature to the desorption of N-related byproducts resulting from the dissociation of DMHy (NH$_{2}$CH$_{3}$, NH$_{3}$ and NH$_{2}$) and/or to the direct desorption of N atoms from the GaP$_{1-x}$N$_{x}$ surface \cite{Onabe1996,Kunert2004,Wakahara2007,Suzuki2007}.

Next, we investigate the influence of the impinging fluxes of DMHy and TBP on the incorporation of N. In principle, one would expect a dependence of \textit{x} on the BEPs of DMHy and TBP given by:
\begin{equation}
  x = \dfrac{s_{1}~BEP_{DMHy}}{s_{1}~BEP_{DMHy}+s_{2}~BEP_{TBP}},
  \label{flux1}
\end{equation}
where s$_{1}$ and s$_{2}$ are coefficients including both the sensitivity of BEP measurements to the particular precursor molecules (caused by their specific ionicity) and the incorporation efficiencies of N and P atoms with respect to the fluxes of DMHy and TBP, respectively \cite{Wildt1999}. Since dilute nitrides are characterized by small \textit{x} values, $s_{1}~BEP_{DMHy}<<s_{2}~BEP_{TBP}$, and Eq.~(\ref{flux1}) can thus be approximated by:
\begin{equation}
  x = \kappa~\dfrac{BEP_{DMHy}}{BEP_{TBP}},
  \label{flux2}
\end{equation}
with $\kappa=s_{1}/s_{2}$ representing the ratio between the N and P incorporation efficiencies with respect to the BEP values of DMHy and TBP. Notice that, according to Eq.~(\ref{eqn:Arrhenius}), $\kappa$ must depend on the growth temperature.

Let us see now whether the incorporation of N as a function of the impinging fluxes is actually well described by Eq.~(\ref{flux2}). Figure~\ref{Figure3}(b) presents the variation of the mean N mole fraction \textit{x} with BEP$_{DMHy}$ at two different growth temperatures, 500 and 550~$^{\circ}$C, using a constant flux of TBP given by BEP$_{TBP}=1.2\times10^{-5}$~Torr. The experimental data show a nearly linear dependence with a temperature dependent slope. The values of $\kappa$ derived from the fits are $(1.54\pm 0.09)\times10^{-2}$ and $(7.72\pm 0.9)\times10^{-3}$ for 500 and 550~$^{\circ}$C, respectively. Consequently, within the investigated temperature range, N incorporation is about two orders of magnitude less efficient than P incorporation with respect to the BEP values of DMHy and TBP. Interestingly, the linear fits does not intercept the x-axis at zero, as predicted by Eq.~(\ref{flux2}). There is, however, a BEP$_{DMHy}$ threshold to overcome in order to incorporate a measurable amount of N. We attribute this threshold to the thermal desorption of N-reactive species and/or N atoms. On these basis, we rewrite the dependence of \textit{x} on the impinging fluxes as:
\begin{equation}
  x = \kappa~\dfrac{BEP_{DMHy}-BEP_{th}}{BEP_{TBP}},
  \label{flux3}
\end{equation}
where $BEP_{th}$ represents the above threshold. The values of $BEP_{th}$ extracted from the fits shown in Fig.~\ref{Figure3}(b) are $7.84\times10^{-6}$ and $9.75\times10^{-6}$~Torr for the series of samples prepared at 500 and 550~$^{\circ}$C, respectively. In consequence, $BEP_{th}$ increases, as expected, with the growth temperature, but the variation is small when compared with the values of\break BEP$_{DMHy}$.

\begin{figure*}
\includegraphics*[width=\textwidth]{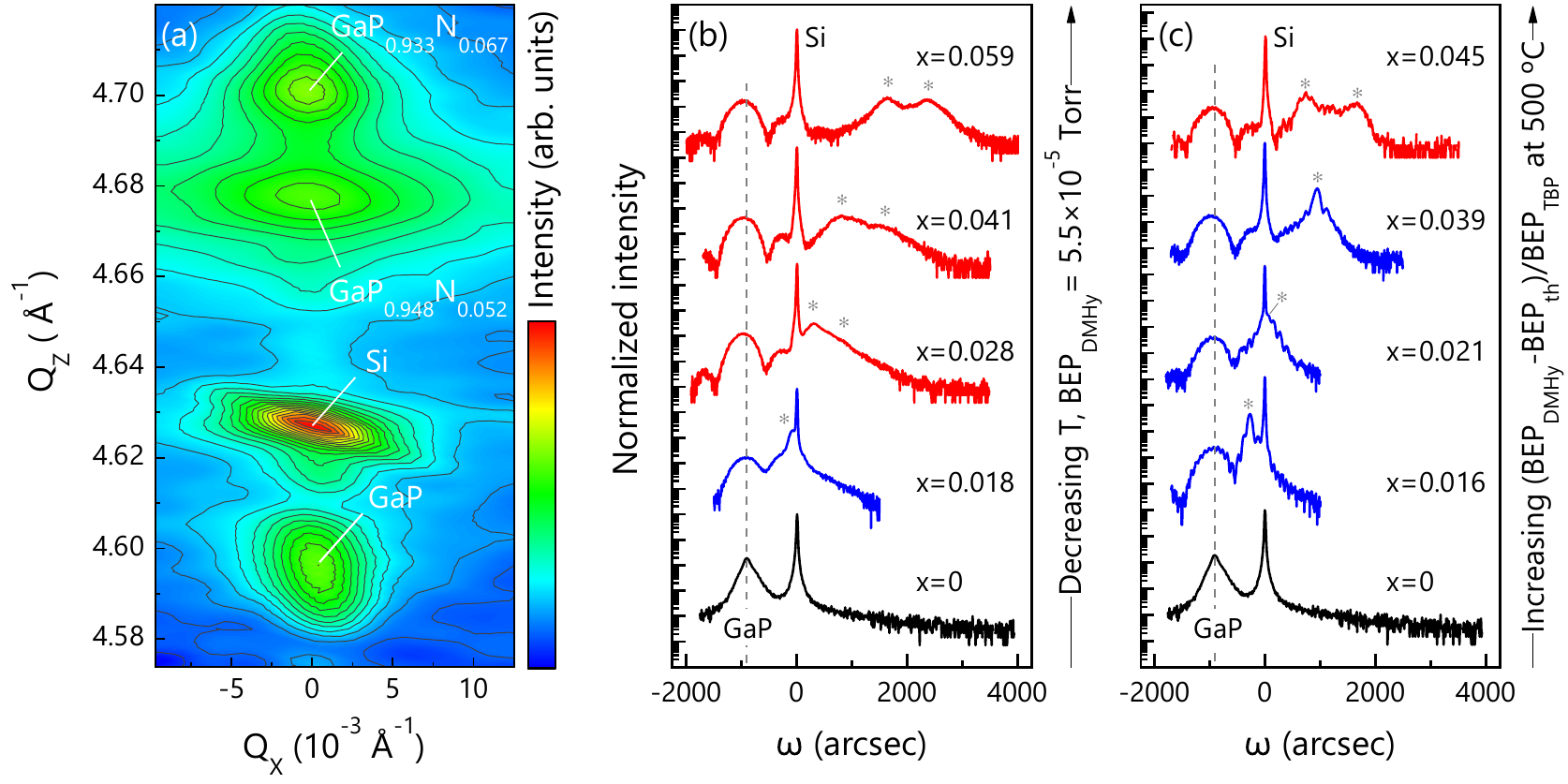}
  \caption{\label{Figure4}(a) Representative RSM around the GaP$_{1-x}$N$_{x}$, GaP and Si symmetrical 004 Bragg reflections of a sample where the GaP$_{1-x}$N$_{x}$ layer exhibits chemical phase separation. For these particular sample, prepared at 500~$^{\circ}$C with BEP$_{DMHy}=5.6\times10^{-5}$~Torr and BEP$_{TBP}=1.2\times10^{-5}$~Torr, we detected two different chemical phases with $x=0.052$ and 0.067. (b and c) $\omega/2\theta$ scans around the GaP$_{1-x}$N$_{x}$, GaP and Si 004 Bragg reflections for two series of samples where $x$ was increased (b) by decreasing the growth temperature from 575 to 500~$^{\circ}$C while keeping constant the supplies of DMHy and TBP (BEP$_{DMHy}=5.5\times10^{-5}$~Torr and BEP$_{TBP}=1.2\times10^{-5}$~Torr), and (c) by rising (BEP$_{DMHy}$-BEP$_{th}$)/BEP$_{TBP}$ at 500~$^{\circ}$C using BEP$_{DMHy}$ values from $2.3\times10^{-5}$ to $5.6\times10^{-5}$~Torr with BEP$_{TBP}=1.2\times10^{-5}$~Torr. The N mole fractions shown in the figures are mean values derived from the combined analysis of symmetrical and asymmetrical RSMs. The symbol * indicates the angular positions of the GaP$_{1-x}$N$_{x}$ 004 Bragg reflections. The scans of samples with and without phase separation are depicted in blue and red colors, respectively. The black scans correspond to a reference $\approx240$~nm thick GaP on Si layer grown at 580~$^{\circ}$C.}
\end{figure*}

To further challenge Eq.~(\ref{flux3}), we plot in Fig.~\ref{Figure3}(c) \textit{x} versus (BEP$_{DMHy}$-BEP$_{th}$)/BEP$_{TBP}$ for the samples shown in Fig.~\ref{Figure3}(b) as well as for three additional samples. These last three samples were grown at 500~$^{\circ}$C with BEP$_{DMHy}=2.8\times10^{-5}$~Torr using different BEP$_{TBP}$ values than the one used to prepare the samples shown in Figs.~\ref{Figure3}(a) and \ref{Figure3}(b), namely, $8.4\times10^{-6}$, $1\times10^{-5}$ and $1.3\times10^{-5}$~Torr. The linear fits included in Fig.~\ref{Figure3}(c) demonstrate that Eq.~(\ref{flux3}) correctly describes all our growth data (the fit for the series prepared at 550~$^{\circ}$C is analogous to the one presented in Fig.~\ref{Figure3}(b), but it is shown here again for the sake of completeness). The $\kappa$ value derived for the samples grown at 500~$^{\circ}$C, $(1.51\pm0.05)\times10^{-2}$, is almost identical to one found before. Therefore, these results reinforce the validity of Eq.~(\ref{flux3}) to describe the dependence of \textit{x} on the impinging fluxes for N mole fractions up to at least 0.06, i.\,e., the maximum \textit{x} value shown in Fig.~\ref{Figure3}(c). 

\subsection{Phase separation}
\label{section1}

The N mole fractions shown in Fig.~\ref{Figure3} are mean values because some GaP$_{1-x}$N$_{x}$ layers are chemically phase separated, as  discussed here. Blue and red data points in Fig.~\ref{Figure3} correspond to single- and phase-separated samples, respectively. Single-phase samples are characterized by exhibiting a unique well-defined GaP$_{1-x}$N$_{x}$ peak in RSMs, as exemplary illustrated in Fig.~\ref{Figure2}. In contrast, phase-separated samples exhibit two GaP$_{1-x}$N$_{x}$ peaks as shown in Fig.~\ref{Figure4}(a) for a sample with a mean N mole fraction of $0.06$ grown at 500~$^{\circ}$C with BEP$_{DMHy}=5.6\times10^{-5}$~Torr and BEP$_{TBP}=1.2\times10^{-5}$~Torr. For this sample, the GaP$_{1-x}$N$_{x}$ peaks detected in the RSM seem to be originated from regions with N mole fractions of $0.052$ and $0.067$. In general, we found that the relative deviation of the chemical composition of the two different phases with respect to the mean chemical composition ($\Delta x/x$) is always below 25~$\%$.  

As can be observed in Figs.~\ref{Figure3}(a) and \ref{Figure3}(c), phase separation occurs as the N mole fraction is increased by either lowering the temperature or increasing the (BEP$_{DMHy}$-BEP$_{th}$)/BEP$_{TBP}$ ratio. The transition between single- and phase-separated GaP$_{1-x}$N$_{x}$ layers as the growth parameters are varied is exemplified in Figs.~\ref{Figure4}(b) and \ref{Figure4}(c), where we present $\omega/2\theta$ HRXRD scans around the symmetric $004$\break Bragg reflections of GaP$_{1-x}$N$_{x}$, GaP and Si for two different series of samples. Blue scans correspond to single-phase samples and red scans to phase-separated samples. As reference, we include in both figures the scan of a $250$~nm thick GaP layer prepared on the same type of substrate as our\break GaP$_{1-x}$N$_{x}$ layers. The 004~GaP peak of the reference sample appears at the same angular position as for the GaP$_{1-x}$N$_{x}$/GaP samples, but it is narrower due to its higher thickness. In Fig.~\ref{Figure4}(b), it can be seen how the GaP$_{1-x}$N$_{x}$ peak, labelled as *, shifts towards larger angles as the N content is increased by lowering the growth temperature while keeping constant the impinging fluxes with BEP$_{DMHy}=5.5\times10^{-5}$~Torr and BEP$_{TBP}$, $1.2\times10^{-5}$~Torr. For these particular fluxes, phase separation starts at some point between $x=0.018$ and $0.028$, as revealed by the split of the GaP$_{1-x}$N$_{x}$ reflection in two different peaks for $x\geq0.028$. Analogously, Fig.~\ref{Figure4}(c) illustrates the onset of phase separation as the N content is increased by raising BEP$_{DMHy}$ at 500~$^{\circ}$C with BEP$_{TBP}=1.2\times10^{-5}$~Torr. Importantly, despite of the similar trend as \textit{x} increases, it is worth to note that in this case it is possible to obtain single-phase samples with comparatively higher N mole fractions, up to a value between 0.039 and 0.045. Therefore, our results reveals that phase separation is not just determined by the N content, but it also depends on the specific growth parameters. Within the framework of this study, the maximum N mole fraction we achieved without phase separation was 0.039.
\begin{figure*}
\includegraphics*[width=0.9\textwidth]{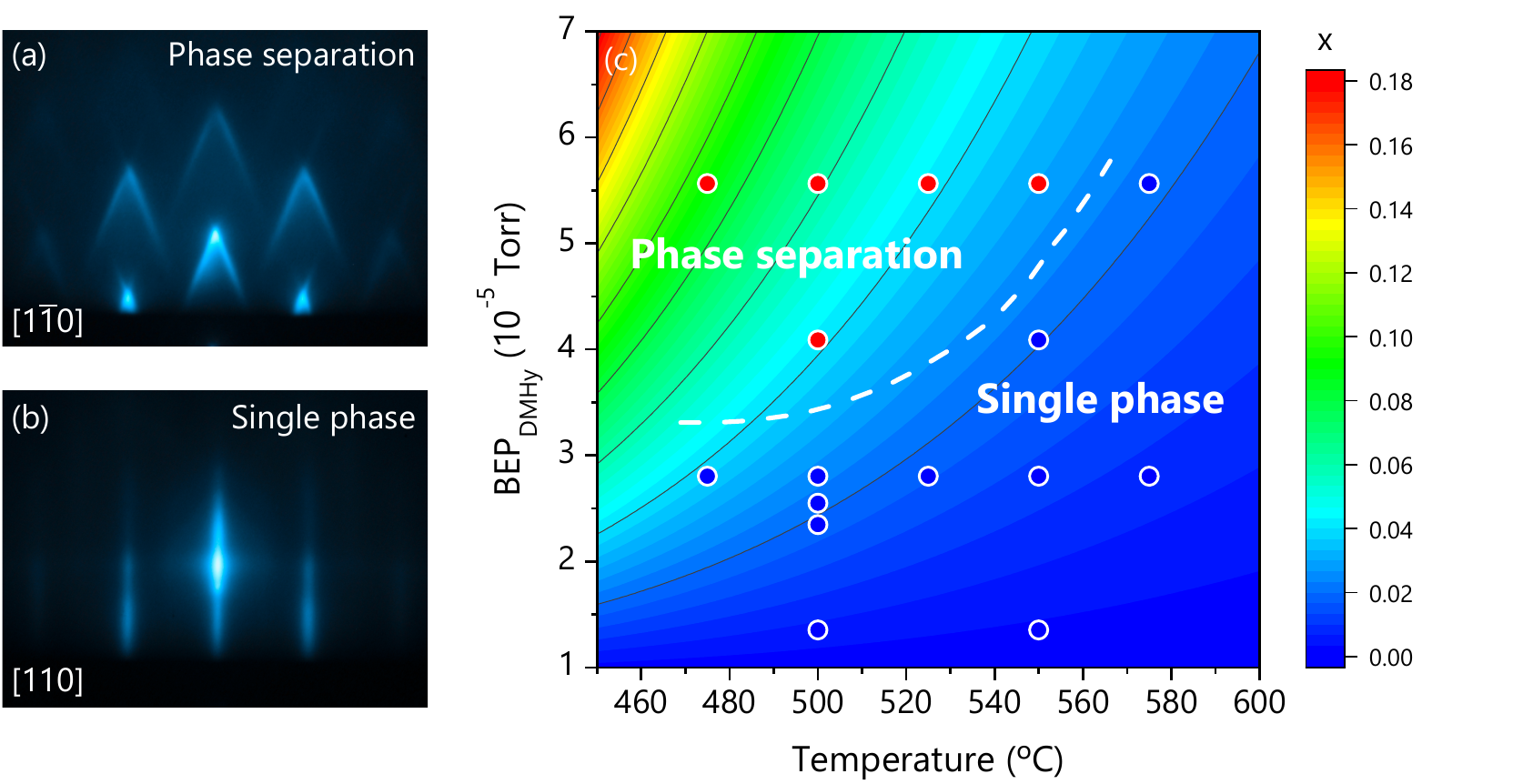}
  \caption{\label{Figure5}(a and b) Characteristic RHEED patterns along the $[1\bar{1}0]$ azimuth upon the growth of (a) phase-separated and (b) single-phase GaP$_{1-x}$N$_{x}$ layers. (c) Growth diagram depicting, according to Eq.~(\ref{model}), the dependence of the mean N mole fraction \textit{x} on the growth temperature and the BEP of DMHy for constant supplies of TBP (BEP$_{TBP}=1.2\times10^{-5}$~Torr) and TEGa (BEP$_{TEGa}=0.3\times10^{-5}$~Torr). The N mole fraction is displayed as a contour plot with a linear scale. The dashed line illustrates the boundary between growth conditions resulting in single-phase and phase-separated GaP$_{1-x}$N$_{x}$ layers. The solid circles indicate the growth conditions of the samples grown within the framework of this study with BEP$_{TBP}=1.2\times10^{-5}$~Torr and BEP$_{TEGa}=0.3\times10^{-5}$~Torr. Blue and red circles represent single-phase and phase-separated samples, respectively. }
\end{figure*}

Last but important, the in-situ monitoring of the growing surface by RHEED revealed an univocal correlation between the growth mode of the GaP$_{1-x}$N$_{x}$ layer and the ex-situ detection of chemical-phase separation by HRXRD. As shown in Fig.~\ref{Figure5}(a), upon the growth of phase-separated GaP$_{1-x}$N$_{x}$ layers, we observed a three-dimensional RHEED pattern along the $[1\bar{1}0]$ azimuth (RHEED azimuths were assigned according to Ref.\citenum{Grassman2009}). The RHEED pattern is composed by chevrons revealing the formation of well-defined facets. According to the mean value of the chevron angles, as measured on raw RHEED screen images, $(46\pm 3)~^{\circ}$, the roughing is likely caused by the formation of $\{113\}$ facets. In contrast, the RHEED pattern of single-phase samples along the $[1\bar{1}0]$ azimuth is streaky, indicating a flat surface morphology [see Fig.~\ref{Figure5}(b)]. These results allow us to conclude that: (i) phase separation can be inferred in situ during growth by RHEED, and (ii) up to at least $x=0.039$, it is possible to synthesize by CBE single-phase GaP$_{1-x}$N$_{x}$ layers with flat surface morphologies.

\subsection{A growth diagram}
All previous results are summarized here in a growth diagram that describes how the chemical and morphological properties of GaP$_{1-x}$N$_{x}$ alloys grown by CBE depend on the growth parameters. Coming back to the control of the alloy composition, by combining Eqs.~(\ref{eqn:Arrhenius}) and (\ref{flux3}), the dependence of the N mole fraction on the growth temperature and the BEP values of the N and P precursors can be written in a compact form as:
\begin{equation}
  x = \beta~\dfrac{BEP_{DMHy}-BEP_{th}}{BEP_{TBP}}~exp~(E_{a}/k_{B}T),
  \label{model}
\end{equation}
where $\beta$ is a constant that might depend only on the impinging flux of the Ga precursor. By taking the mean values of $E_{a}$ and BEP$_{th}$ as 0.79~eV and $1.2\times10^{-5}$~Torr, we derived from the values of \textit{A} and $\kappa$ reported in section~\ref{section3.1} a $\beta$ value of $1.15\times10^{-7}$ for the Ga flux used in this study, i.\,e., 0.22~ML/s. 

Figure~\ref{Figure5}(c) depicts a growth diagram, which visualizes according to Eq.~(\ref{model}) the impact of the growth temperature and BEP$_{DMHy}$ on the incorporation of N into GaP$_{1-x}$N$_{x}$. This diagram quantitatively holds for BEP$_{TBP}=1.2\times10^{-5}$ \break Torr and the Ga flux used in this work. Nevertheless, analogous diagrams could be constructed for arbitrary values of BEP$_{TBP}$ using Eq.~(\ref{model}). Regarding the role of the actual Ga flux, it could affect the specific values of \textit{x}, but we expect a comparable qualitative dependence of the N mole fraction on the growth temperature and BEP$_{DMHy}$. 

Besides the variation of \textit{x} with the growth temperature and BEP$_{DMHy}$, we also indicate in the growth diagram both the growth conditions of the samples grown in this study with BEP$_{TBP}=1.2\times10^{-5}$~Torr (solid symbols), and the approximate boundary between the growth parameters resulting in single-phase and phase-separated GaP$_{1-x}$N$_{x}$ layers (dashed line), as determined by the analysis of the samples by HRXRD. This boundary also applies to distinguish between growth conditions leading to flat and rough surface morphologies, as single-phase and phase-separated samples exhibit two- and three-dimensional RHEED patterns, respectively, as discussed above. The accuracy of our growth diagram was tested by comparing the N mole fractions shown in there with the actual ones measured by HRXRD for the samples prepared using the parameters indicated by the solid symbols shown in Fig.~\ref{Figure5}(c). The relative mean deviation between the expected and actual \textit{x} values ($\Delta x/x$) is $\approx10\%$. The diagram thus provides a rather satisfactory description of \textit{x} when considering all experimental uncertainties. Hence, Eq.~(\ref{model}) can be used to predict, with the aforementioned precision, the chemical composition of GaP$_{1-x}$N$_{x}$ layers grown under unexplored growth conditions, at least for \textit{x} values within the range analyzed here, i.\,e., $0<x<0.09$. 

To conclude, the growth diagram also evidences that identical chemical compositions can be achieved by combining different growth temperatures and BEP$_{DMHy}$ values, as these two parameters play opposite roles for the incorporation of N. It is also important to note that the phase-separated growth regime tends to extend towards lower values of \textit{x} as the\break growth temperature is increased. Because of this effect and the exponential temperature dependence of \textit{x} on the growth temperature, we recommend using moderate growth temperatures to prevent phase separation as well as to minimize the impact of possible temperature deviations on the final alloy composition. Nevertheless, it remains to be seen the impact of the growth temperature on point defects formation.

\section{Summary and conclusions}
We have comprehensively investigated the incorporation of N into GaP$_{1-x}$N$_{x}$ layers grown on $(001)$-oriented GaP-on-Si substrates by CBE using as gas sources DMHy, TBP and TEGa. According to our study, the N mole fraction: (i) exhibits an Arrhenius-like temperature dependence characterized by an apparent activation energy of $(0.79\pm 0.05)$~eV, (ii) increases lineraly with the flux of DMHy, and (iii) is inversely proportional to the flux of TBP. The results are summarized in an empirical universal equation that can be used to predict the N mole fraction within the compositional range analyzed in this study, $0<x<0.09$, with a relative degree of uncertainty of $\approx 10\%$ over the targeted value. All results are further illustrated in a growth diagram, which does not only visualize the dependence of \textit{x} on the growth temperature and the flux of DMHy for given fluxes of TBP and TEGa, but also emphasizes under which growth conditions it is possible to obtain single-phase and flat GaP$_{1-x}$N$_{x}$ layers. Importantly, although as a general trend chemical phase separation is triggered as the N mole fraction is increased, the threshold for the onset of phase separation is not just determined by \textit{x}, it depends as well on the specific growth parameters. The maximum value of \textit{x} we achieved for single-phase and flat GaP$_{1-x}$N$_{x}$ layers was 0.039. This value is well above 0.021, i.\,e., the required N mole fraction for the synthesis of GaP$_{1-x}$N$_{x}$ layers lattice matched to Si$(001)$. This work thus evidences the feasibility of CBE for the monolithic integration of pseudomorphic GaP$_{1-x}$N$_{x}$-based optoelectronic devices on Si.

\section*{Data Availability Statement}

The data that support the findings of this study are available
from the corresponding author upon reasonable request.

\begin{acknowledgments}
This work was supported by the former Ministerio de Ciencia, Innovación y Universidades and the current Ministerio de Ciencia e Innovación under Projects No. TEC2016-78433-R and PID2020-114280RB-I00, respectively. Additionally, S. Fernández-Garrido acknowledges the financial support received through the program Ramón y Cajal (co-financed by the European Social Fund) under Grant No. RYC-2016-19509 from Ministerio de Ciencia, Innovación y Universidades. We also thank to T. Vallés for his technical support and assistance. 
\end{acknowledgments}

\section*{References}
\nocite{*}
\bibliographystyle{unsrt}
\bibliography{aipsamp}

\providecommand{\noopsort}[1]{}\providecommand{\singleletter}[1]{#1}%
\begin{thebibliography}{10}

\bibitem{Shan2004}
W~Shan, K~M Yu, W~Walukiewicz, J~Wu, J~W Ager, and E~E Haller.
\newblock {Band anticrossing in dilute nitrides}.
\newblock {\em J. Phys.: Condens. Matter}, 16:S3355, 2004.

\bibitem{Chamings2009}
J~Chamings, S~Ahmed, A~R Adams, S~J Sweeney, V~A Odnoblyudov, C~W Tu, B~Kunert,
  and W~Stolz.
\newblock {Band anti-crossing and carrier recombination in dilute nitride
  phosphide based lasers and light emitting diodes}.
\newblock {\em Phys. Status Solidi B}, 246:527, 2009.

\bibitem{Xin2000a}
H~P Xin, R~J Welty, and C~W Tu.
\newblock {{GaN$_{0.011}$P$_{0.989}$} red light-emitting diodes directly grown
  on GaP substrates}.
\newblock {\em Appl. Phys. Lett.}, 77:1946, 2000.

\bibitem{Cornet2011}
C~Cornet, C~Robert, T~N Thanh, W~Guo, A~Bondi, G~Elias, A~L{\'{e}}toublon,
  S~Richard, J~P Burin, M~Perrin, J~M Jancu, O~Durand, J~Even, S~Loualiche,
  H~Folliot, N~Bertru, A~Ponchet, and A~L Corre.
\newblock {Carrier injection in {GaAsPN}/{GaPN} quantum wells on {Silicon}}.
\newblock {\em IPRM 2011 - 23rd International Conference on Indium Phosphide
  and Related Materials}, page~1, 2011.

\bibitem{Robert2012}
C~Robert, A~Bondi, T~N Thanh, J~Even, C~Cornet, O~Durand, and J~P Burin.
\newblock {Room temperature operation of GaAsPN/GaPN quantum well based
  light-emitting diodes: Effect of the incorporation of nitrogen}.
\newblock 251110:10, 2012.

\bibitem{Jain2014}
N~Jain and M~K Hudait.
\newblock {{III}–{V} {Multijunction} {Solar} {Cell} {Integration} with
  {Silicon}: {Present} {Status}, {Challenges} and {Future} {Outlook}}.
\newblock {\em Energy Harvesting and Systems}, 1, 2014.

\bibitem{Durand2014}
O~Durand, S~Almosni, Y~{Ping Wang}, C~Cornet, A~L{\'{e}}toublon, C~Robert,
  C~Levallois, L~Pedesseau, A~Rolland, J~Even, J~M Jancu, N~Bertru, A~{Le
  Corre}, F~Mandorlo, M~Lemiti, P~Rale, L~Lombez, J.-F. Guillemoles, S~Laribi,
  A~Ponchet, and J~Stodolna.
\newblock {Monolithic {Integration} of {Diluted}-{Nitride} {III}–{V}-{N}
  {Compounds} on {Silicon} {Substrates}: {Toward} the {III}–{V}/{Si}
  {Concentrated} {Photovoltaics}}.
\newblock {\em Energy Harvesting and Systems}, 1, 2014.

\bibitem{Sukrittanon2015}
S~Sukrittanon, R~Liu, Y~G Ro, J~L Pan, K~L Jungjohann, C~W Tu, and S~A Dayeh.
\newblock {Enhanced conversion efficiency in wide-bandgap {GaNP} solar cells}.
\newblock {\em Appl. Phys. Lett.}, 107:153901, 2015.

\bibitem{Kharel2018}
K~Kharel and A~Freundlich.
\newblock {Band structure and absorption properties of
  ({Ga},{In})/({P},{As},{N}) symmetric and asymmetric quantum wells and
  super-lattice structures: {Towards} lattice-matched {III}-{V}/{Si} tandem}.
\newblock {\em J. Appl. Phys.}, 124:95104, 2018.

\bibitem{BenSaddik2019}
K~{Ben Saddik}, A~F Bra{\~{n}}a, N~L{\'{o}}pez, W~Walukiewicz, and B~J
  Garc{\'{i}}a.
\newblock {Growth of GaP$_{1-x-y}$As$_{y}$N$_{x}$ on Si substrates by chemical
  beam epitaxy}.
\newblock {\em J. Appl. Phys.}, 126:105704, 2019.

\bibitem{Tsao2014}
J~Y Tsao, M~H Crawford, M~E Coltrin, A~J Fischer, D~D Koleske, G~S Subramania,
  G~T Wang, J~J Wierer, and R~F Karlicek.
\newblock {Solid-{State} {Lighting}: {Toward} {Smart} and {Ultra}-efficient
  {Solid}-{State} {Lighting}}.
\newblock {\em Advanced Optical Materials}, 2:803, 2014.

\bibitem{Yonezu2002}
H~Yonezu.
\newblock {Control of structural defects in group {III} – {V} – {N} alloys
  grown on {Si}}.
\newblock {\em Semicond. Sci. Technol.}, 17:762, 2002.

\bibitem{Almosni2013}
S~Almosni, C~Robert, T~{Nguyen T.}, C~Cornet, A~L{\'{e}}toublon, T~Quinci,
  C~Levallois, M~Perrin, J~Kuyyalil, L~Pedesseau, A~Balocchi, P~Barate, J~Even,
  J~M Jancu, N~Bertru, X~Marie, O~Durand, and A~{Le Corre}.
\newblock {Evaluation of {InGaPN} and {GaAsPN} materials lattice-matched to
  {Si} for multi-junction solar cells}.
\newblock {\em J. Appl. Phys.}, 113:123509, 2013.

\bibitem{Tilli2014}
J.-M. Tilli, H~Jussila, K~M Yu, T~Huhtio, and M~Sopanen.
\newblock {Composition determination of quaternary {GaAsPN} layers from single
  {X}-ray diffraction measurement of quasi-forbidden (002) reflection}.
\newblock {\em J. Appl. Phys.}, 115:203102, 2014.

\bibitem{Yamane2018}
K~Yamane, S~Mugikura, S~Tanaka, M~Goto, H~Sekiguchi, H~Okada, and A~Wakahara.
\newblock {Impact of temperature and nitrogen composition on the growth of
  {GaAsPN} alloys}.
\newblock {\em J. Cryst. Growth}, 486:24, 2018.

\bibitem{Balanta2020}
M~A~G Balanta, P~B~A de~Oliveira, H~Albalawi, Y~Galv{\~{a}}o Gobato, H~V~A
  Galeti, A~D Rodrigues, M~Henini, S~Almosni, C~Robert, A~Balocchi,
  Y~L{\'{e}}ger, H~Carr{\`{e}}re, M~Bahri, G~Patriarche, X~Marie, and C~Cornet.
\newblock {Effects of nitrogen incorporation and thermal annealing on the
  optical and spin properties of {GaPN} dilute nitride alloys}.
\newblock {\em J. Alloys Compd.}, 814:152233, 2020.

\bibitem{Nagarajan2013}
S~Nagarajan, H~Jussila, J~Lemettinen, K~Banerjee, M~Sopanen, and H~Lipsanen.
\newblock {Strain-compensated {GaPN}/{GaP} heterostructure on (001) silicon
  substrates for intermediate band solar cells}.
\newblock {\em J. Phys. D: Appl. Phys.}, 46:165103, 2013.

\bibitem{Li2017}
Q~Li and Kei~M Lau.
\newblock {Epitaxial growth of highly mismatched {III}-{V} materials on (001)
  silicon for electronics and optoelectronics}.
\newblock {\em Prog. Cryst. Growth Charact. Mater.}, 63:105, 2017.

\bibitem{Houng1992}
Y.-M. Houng.
\newblock {Chemical beam epitaxy}.
\newblock {\em Crit. Rev. Solid State Mater. Sci.}, 17:277, 1992.

\bibitem{MiguelSanchez2004}
J~Miguel-S{\'{a}}nchez, A~Guzm{\'{a}}n, and E~Mu{\~{n}}oz.
\newblock {Role of {N} ions in the optical and morphological properties of
  {InGaAsN} quantum wells for 1.3–1.5 $\mu$m applications}.
\newblock {\em Appl. Phys. Lett.}, 85:1940, 2004.

\bibitem{Utsumi2003}
A~Utsumi, H~Yonezu, Y~Furukawa, K~Momose, and K~Kuroki.
\newblock {Increase in luminescence efficiency of {GaPN} layers by thermal
  annealing}.
\newblock {\em Phys. Status Solidi C}, 0:2741, 2003.

\bibitem{Moon2004}
S~Y Moon, A~Utsumi, H~Yonezu, Y~Furukawa, T~Ikeda, and A~Wakahara.
\newblock {Ga{P}{N}-{G}a{P} double heterostructure light emitting diode grown
  on GaP substrate by solid-source molecular beam epitaxy}.
\newblock {\em Phys. Status Solidi A}, 201:2695, 2004.

\bibitem{Li1996}
N~Y Li, W~S Wong, D~H Tomich, H~K Dong, J~S Solomon, J~T Grant, and C~W Tu.
\newblock {Growth study of chemical beam epitaxy of {Ga}{N}$_{x}${P}$_{1-x}$
  using {N}{H}$_{3}$ and tertiarybutylphosphine}.
\newblock {\em J. Cryst. Growth}, 164:180, 1996.

\bibitem{Suemune2002}
I~Suemune, T~Shimozawa, K~Uesugi, H~Kumano, H~Machida, and N~Shimoyama.
\newblock {Metalorganic {Molecular}-{Beam} {Epitaxial} {Growth} and {Optical}
  {Properties} of {Er}-{Doped} {GaNP}}.
\newblock {\em Jpn. J. Appl. Phys.}, 41:1030, 2002.

\bibitem{AitLhouss1994}
M~Ait-Lhouss, J~L Casta{\~{n}}o, and J~Piqueras.
\newblock {Chemical beam epitaxial growth of {GaAs} from tertiarybutylarsine
  and triethylgallium precursors}.
\newblock {\em Mater. Sci. Eng. B}, 28:155, 1994.

\bibitem{Roesener2013}
Tobias Roesener, Vera Klinger, Christoph Weuffen, David Lackner, and Frank
  Dimroth.
\newblock {Determination of heteroepitaxial layer relaxation at growth
  temperature from room temperature {X}-ray reciprocal space maps}.
\newblock {\em J. Cryst. Growth}, 368:21, 2013.

\bibitem{Chauveau2003}
J.-M. Chauveau, Y~Androussi, A~Lefebvre, J~{Di Persio}, and Y~Cordier.
\newblock {Indium content measurements in metamorphic high electron mobility
  transistor structures by combination of x-ray reciprocal space mapping and
  transmission electron microscopy}.
\newblock {\em J. Appl. Phys.}, 93:4219, 2003.

\bibitem{Takagi2010}
Yasufumi Takagi, Yuzo Furukawa, Akihiro Wakahara, and Hirofumi Kan.
\newblock {Lattice relaxation process and crystallographic tilt in {GaP} layers
  grown on misoriented {Si}(001) substrates by metalorganic vapor phase
  epitaxy}.
\newblock {\em J. Appl. Phys.}, 107:63506, 2010.

\bibitem{Vurgaftman2001}
I~Vurgaftman, J~R Meyer, and L~R Ram-Mohan.
\newblock {Band parameters for {III}–{V} compound semiconductors and their
  alloys}.
\newblock {\em J. Appl. Phys.}, 89(11):5815, 2001.

\bibitem{Bi1996}
W~G Bi and C~W Tu.
\newblock {N incorporation in {GaP} and band gap bowing of
  {GaN$_{x}$P$_{1-x}$}}.
\newblock {\em Appl. Phys. Lett.}, 69:3710, 1996.

\bibitem{Kunert2004}
B~Kunert, J~Koch, T~Torunski, K~Volz, and W~Stolz.
\newblock {{MOVPE} growth experiments of the novel ({GaIn})({NP})/{GaP}
  material system}.
\newblock {\em J. Cryst. Growth}, 272:753, 2004.

\bibitem{Wakahara2007}
A~Wakahara, Y~Furukawa, S~Itoh, S~Hatakenaka, and H~Yonezu.
\newblock {Growth and characterization of {GaPN} by {OMVPE}}.
\newblock {\em J. Cryst. Growth}, 300:182, 2007.

\bibitem{Onabe1996}
K~Onabe.
\newblock {{MOVPE} {Growth} and {Optical} {Characterization} of {GaPN}
  {Metastable} {Alloy} {Semiconductor}}.
\newblock {\em MRS Proc.}, 449:23, 1996.

\bibitem{Suzuki2007}
H~Suzuki, K~Nishimura, H~S Lee, Y~Ohshita, N~Kojima, and M~Yamaguchi.
\newblock {Carbon incorporation process in {GaAsN} films grown by chemical beam
  epitaxy using {MMHy} or {DMHy} as the {N} source}.
\newblock {\em Thin Solid Films}, 515:5008, 2007.

\bibitem{Wildt1999}
D~Wildt, B~J Garc{\'{i}}a, J~L Casta{\~{n}}o, and J~Piqueras.
\newblock {Phosphorus and arsenic incorporation during chemical beam epitaxial
  growth of strained GaAs$_{1-x}$P$_{x}$ layers on GaAs(100) substrates}.
\newblock {\em Proceedings of the 10$^{th}$ Conference on Semiconducting and
  Insulating Materials (SIMC-X) (Cat. No.98CH36159), IEEE}, page~15, 1999.

\bibitem{Grassman2009}
T.~J. Grassman, M.~R. Brenner, S.~Rajagopalan, R.~Unocic, R.~Dehoff, M.~Mills,
  H.~Fraser, and S.~A. Ringel.
\newblock {Control and elimination of nucleation-related defects in GaP/Si(001)
  heteroepitaxy}.
\newblock {\em App. Phys. Lett.}, 94(23):232106, 2009.

\end{thebibliography}

\end{document}